\documentclass[preprint, 5p, times]{elsarticle}
\journal{Health Policy and Technology}

\usepackage{hyperref} % PDF links
\usepackage{subcaption}

\usepackage{caption}
\usepackage{setspace}
\usepackage{natbib}
\usepackage{csquotes}
\usepackage[UKenglish]{babel}

\begin{document}
%
% paper title
% Titles are generally capitalized except for words such as a, an, and, as,
% at, but, by, for, in, nor, of, on, or, the, to and up, which are usually
% not capitalized unless they are the first or last word of the title.
% Linebreaks \\ can be used within to get better formatting as desired.
% Do not put math or special symbols in the title.
\begin{frontmatter}

\title{Smart Home Goal Feature Model - A guide to support Smart Homes for Ageing in Place}
%
%
% author names and IEEE memberships
% note positions of commas and nonbreaking spaces ( ~ ) LaTeX will not break
% a structure at a ~ so this keeps an author's name from being broken across
% two lines.
% use \thanks{} to gain access to the first footnote area
% a separate \thanks must be used for each paragraph as LaTeX2e's \thanks
% was not built to handle multiple paragraphs
%

% \author{Michael~Shell,~\IEEEmembership{Member,~IEEE,}
%         John~Doe,~\IEEEmembership{Fellow,~OSA,}
%         and~Jane~Doe,~\IEEEmembership{Life~Fellow,~IEEE}% <-this % stops a space
%\author{Priya~Rani,~Rena~Logothetis,~Mahdi~Babaei,~Rajesh~Vasa,~Kon~Mouzakis%
%\author{Priya~Rani$^{1}$,~Rena~Logothetis$^{1}$,~Mahdi~Babaei$^{1}$,~Rajesh~Vasa$^{1}$,~Kon~Mouzakis$^{1}$}%

%\thanks{P. Rani, R. Logothetis, M. Babaei, R. Vasa and K. Mouzakis are with Applied Artificial Intelligence Institute, Deakin University, Burwood, Australia. Email: priya.rani@deakin.edu.au}}%
%\thanks{$^{2}$N. Bowes and T. Soderlund are with Uniting Agewell, Victoria, Australia}%

\author[1]{Irini Logothetis}\corref{cor1}
\ead{rena.logothetis@deakin.edu.au}
\author[2]{Priya Rani}
\ead{priya.rani@rmit.edu.au}
\author[1]{Shangeetha Sivasothy}
\ead{s.sivasothy@deakin.edu.au}
\author[1]{Rajesh Vasa}
\ead{rajesh.vasa@deakin.edu.au}
\author[1]{Kon Mouzakis}
\ead{kon.mouzakis@deakin.edu.au}
\cortext[cor1]{Corresponding author}

\affiliation[1]{organization={Applied Artificial Intelligence Institute},
addressline={Deakin University},
postcode={3125},
city={Burwood},
country={Australia}}

\affiliation[2]{organization={School of Engineering},
addressline={RMIT University},
postcode={3125},
city={Burwood},
country={Australia}}

\begin{abstract}
Smart technologies are significant in supporting ageing in place for elderly. Leveraging Artificial Intelligence (AI) and Machine Learning (ML), it provides peace of mind, enabling the elderly to continue living independently. Elderly use smart technologies for entertainment and social interactions, this can be extended to provide safety and monitor health and environmental conditions, detect emergencies and notify informal and formal caregivers when care is needed. This paper provides an overview of the smart home technologies commercially available to support ageing in place, the advantages and challenges of smart home technologies, and their usability from elderly's perspective. Synthesizing prior knowledge, we created a structured Smart Home Goal Feature Model (SHGFM) to resolve heuristic approaches used by the Subject Matter Experts (SMEs) at aged care facilities and healthcare researchers in adapting smart homes. The SHGFM provides SMEs the ability to (i) establish goals and (ii) identify features to set up strategies to design, develop and deploy smart homes for the elderly based on personalised needs. Our model provides guidance to healthcare researchers and aged care industries to set up smart homes based on the needs of elderly, by defining a set of goals at different levels mapped to a different set of features.
\end{abstract}

% Note that keywords are not normally used for peerreview papers.
%\begin{IEEEkeywords}
%Smart home, Ageing in place, Goal feature model
%\end{IEEEkeywords}


\end{frontmatter}

% For peer review papers, you can put extra information on the cover
% page as needed:
% \ifCLASSOPTIONpeerreview
% \begin{center} \bfseries EDICS Category: 3-BBND \end{center}
% \fi
%
% For peerreview papers, this IEEEtran command inserts a page break and
% creates the second title. It will be ignored for other modes.
%\IEEEpeerreviewmaketitle

\section{Introduction}
% The very first letter is a 2 line initial drop letter followed
% by the rest of the first word in caps.
% 
% form to use if the first word consists of a single letter:
% \IEEEPARstart{A}{demo} file is ....
% 
% form to use if you need the single drop letter followed by
% normal text (unknown if ever used by the IEEE):
% \IEEEPARstart{A}{}demo file is ....
% 
% Some journals put the first two words in caps:
% \IEEEPARstart{T}{his demo} file is ....
% 
% Here we have the typical use of a "T" for an initial drop letter
% and "HIS" in caps to complete the first word.
%\IEEEPARstart{I}{n} 
In the last two decades, life expectancy has increased worldwide. According to the UN Department of Economic and Social Affairs Population Division, the worldwide life expectancy is expected to increase from 46–89 years to 66–93 years in the 21st century \cite{UNations2004World2300}. During 2100 to 2300, people aged over 65 will increase from 24\% to 32\%, and those over 80 will double. Thus, we expect a large ageing population with functional limitations and co-morbidities, burdening an already stretched healthcare sector. An economic need for population-based interventions to support \textquote{ageing in place} has been recognized \cite{Bottazzi2006Context-awarePeople, Freitas2015CombiningPeople} necessitating flexible services accessible to elderly living in the community and health monitoring and management systems to predict, prevent and manage their health improving their mental well-being.

Currently, ageing facilities and carers heuristically adapt smart homes to monitor elderly. The advancement of Artificial Intelligence and Machine Learning has increased the features available in commercially available devices where we can monitor vitals, such as heart rate and blood pressure, or detect falls. These devices can be synced to services such as smartphone apps or emails to communicate with their carers informing them of their health status or fall incidents. In practice, these devices are placed ad-hoc in homes for monitoring. 

Tailored devices provide support where data captured is translated into information that can be shared with health professionals remotely, supporting remote monitoring. Their clinical experience lead to actionable decisions to intervene when elderly are at risk, providing elderly a sense of independent living at home with peace of mind that they are managed. For example, sending an ambulance to the residence. Adapting smart homes can improve healthcare services for an `ageing in place' community. The challenges in setting up smart homes faced by Subject Matter Experts (SMEs), such as healthcare researchers and aged care industries, are threefold (i) needs assessment of elderly, (ii) identifying features that map to those needs and in the field of Gerontology and (iii) establishing goals. 

The scope of this work is to review research on smart homes for elderly and synthesise this knowledge into a structured Smart Home Goal Feature Model (SHGFM) to support SMEs address (i)-(iii). This supports SMEs to set strategic goals in designing and deploying smart homes and identify trade-offs across various maturity levels such as resources, costs, development and support. Our model incorporates a layered feature maturity assessment strategy for SMEs to achieve the above goals.

%This paper covers the use of smart technology for the elderly and presents a structured SHGFM that can be considered for setting up smart homes for the elderly. 
This paper presents research of smart home technologies to support \textquote{ageing in place} \autoref{sec:smarttech}, \autoref{sec:perspective} providing an overview and case studies from an end user perspective. The advantages and research challenges for implementing smart home technologies for elderly have been summarised in \autoref{sec:advantages} and \autoref{sec:challenges} respectively. \autoref{sec:shgfm} presents our SHGFM. We conclude by discussing future perspectives in \autoref{sec:conclusion}. 

% You must have at least 2 lines in the paragraph with the drop letter
% (should never be an issue)

% \hfill mds
 
% \hfill August 26, 2015

\section{Smart Technology and IoT for ageing in place} \label{sec:smarttech}

\begin{figure}[b!]
\centering
\includegraphics[width=0.9\linewidth]{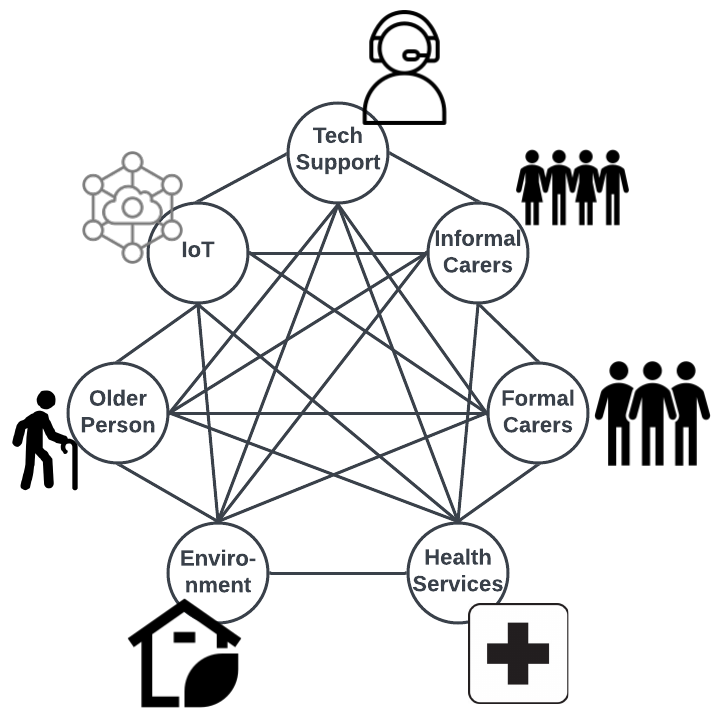}
\caption{The network of smart home technology to support ageing in place}
\label{fig_sim1}
\end{figure}

Research indicates that elderly prefer living independently at home to an aged care facility \cite{Freitas2015CombiningPeople, FahimDailySmartphone}. \textquote{Ageing in place} supports this irrespective of their health and mobility deterioration \cite{Carnemolla2018AgeingIntersect, Pal2018Internet-of-ThingsPerspective}. It involves a collaborative effort from elderly, informal caregivers such as partners, family and friends and their formal care networks including medical practitioners, aged care workers and therapists extending to technical providers. Smart homes incorporating home-based consumer health technologies that can connect to remote healthcare services enhances \textquote{ageing in place}. \autoref{fig_sim1} extends on \cite{Carnemolla2018AgeingIntersect} presenting the smart home technology environment and important stakeholders to include technical support and health services. 

%\begin{figure}[b!]
%\centering
%\includegraphics{Figure1.png}
%\centering
%\caption{The network of smart home technology to support ageing in place %\cite{Carnemolla2018AgeingIntersect}}
%\label{fig_sim1}
%\end{figure}

The  Internet of Things (IoT) has advanced healthcare systems to support \textquote{ageing in place} \cite{Majumder2017SmartChallenges, Wang2014AnNetworks}. IoT defined as a network of objects that measure, collect, process and share data and transmit feedback used for decision-making \cite{Pal2018Internet-of-ThingsPerspective, Majumder2017SmartChallenges}, are integrated into smart homes for automatic and remote monitoring, providing safety and security to elderly.

There are four main functions of smart home systems: 
\begin{enumerate}
\item Monitoring health and environmental conditions, 
\item Detecting anomalies and emergencies, 
\item Notifying informal and formal caregivers of changes in the health status of elderly, and
\item Providing entertainment and social interactions. 
\end{enumerate}

There are five basic types of smart home systems developed for elderly:
\begin{enumerate}
\item Automated emergency call systems,
\item Activity and fall detection systems,
\item Vital signs monitoring systems, 
\item Cognitive assistance reminding systems, and
\item Social and entertainment systems. 
\end{enumerate}

\subsection{Automated emergency call systems}

Automated emergency call systems leverage wearable devices and smartphones to communicate emergencies to caregivers (informal and formal). \cite{Bottazzi2006Context-awarePeople} uses a wearable ECG PocketView Holter to monitor arrhythmia disorders alerting caregivers of any detected anomalies in the physiological data. \cite{Freitas2015CombiningPeople} is a hazard prediction system that collects and analyses motion data from a webcam, sending hazard notifications to a users smartphone and calls predefined numbers for assistance . \cite{Choudhury2015DesignSystem} presents a smartphone-based security and alarm system where the elderly can press a panic button at any location and the emergency contact receives an alert and GPS location. 

\subsection{Automated fall and activity detection systems}

Falls are the leading cause of injuries among elderly. Video-based monitoring systems can track activities to detect falls and alert caregivers \cite{Zhou2009AElders, NiRGBD-HuDaAct:Recognition} by analysing gait activities; however, elderly consider this an invasion of privacy \cite{Zhou2009AElders}. HomeGuardian \cite{HomeGuardianAIPtyLtd2021HomeGuardian} uses patent technology to process image data in real-time locally transmitting only fall alerts. No images or videos are taken, stored, sent or shared by the device – even in the event of incident detection, thus maintaining privacy \cite{HomeGuardianAIPtyLtd2021HomeGuardian}. Wearable devices that detect falls within or outside the house \cite{HoaBinhLe2007HealthElders, Dawadi2013AutomatedTasks} have poor adoption by elderly. A smartphone-based fall detection system can detect falls by measuring and analysing user movement \cite{AbbateASystem}. Hybrid systems use wearable devices to detect falls and triggers audio and video capture to verify it \cite{Wang2014AnNetworks, ZhangHONEY:System}. Garments can be used to detect activities and falls among the elderly unintrusively \cite{MandlerInternet1, KompatsiarisIE13Environments} by embedding motion sensors into garments such as armbands, t-shirts and socks. These are limited by user adherence and charging.

%\begin{figure}[b!]
%\centering
%\includegraphics{Figure2.png}
%\centering
%\caption{The workflow of a remote fall detection system \cite{Majumder2017SmartChallenges}}
%\label{fig_sim2}
%\end{figure}

\subsection{Vital signs monitoring systems}

Vital signs and physiological parameters are measured using wearable and non-wearable sensors \cite{logothetis2019embroidered, logothetis2020evaluating, logothetis2018optimum, logothetis2021comparison}. Multi-parameter monitoring systems in remote monitoring of elderly \cite{Shivakumar2014DesignTechnology, Deen2015InformationHome, Adib2015SmartRate} capture temperature, heart rate (HR), heart rate variability (HRV), blood pressure (BP), respiration rate (RR), electrocardiogram (ECG), electromyogram (EMG) and oxygen saturation (SpO2). For elderly less mobile, physiological data is collected through smart furniture (smart beds and chairs) detecting the number of apneic episodes, depth of sleep and early symptoms of sleep disorders through monitoring sleep patterns and heart rate \cite{HolzingerHuman-ComputerBeyond, Barsocchi2017AnUnderstanding, Nam2016SleepSensor}.

\subsection{Cognitive assistance systems}

Cognitive assistive technologies help elderly in their daily routines through automated alerts at pre-scheduled times reminding them and caregivers to keep hydrated, eat and take medications \cite{FahimDailySmartphone, Moshnyaga2017ADementia}. In addition, task instruction technologies prove useful for elderly with memory and cognitive decline, such as audio or video instructions on using  appliances \cite{Demiris2008IMIAApplications}. Wedjat reminds elderly to take their medicines \cite{Zao2010SmartMonitor} providing intake directions while capturing quality real-time data such as ``taken'' and ``missed'' medication.

\subsection{Social and entertainment systems}

Technologies facilitating entertainment and social interactions combat social isolation. These include televisions, gaming technology and robots, and social interactive multimedia, such as messages, audio and video calls with friends and family and virtual participation in group activities or events appear on smart devices \cite{Demiris2008IMIAApplications, Kon2017EvolutionElderly, Fernando2016EXAMININGELDERLY}. Through AI, smart televisions and gaming technology can be personalised providing entertainment to elderly based on their wants and needs, recommending news, music, digital gaming and digital drawing. Digital games provide physical exercise and improve cognitive functions \cite{Kon2017EvolutionElderly}. \cite{Kon2017EvolutionElderly} presents social robots that interact giving advice on activity level, weather, morning dialog, and going out and coming home dialog .

%\begin{figure}[hbt!]
%\centering
%\includegraphics[width=0.9\linewidth]{Figure3.png}
%\centering
%\caption{Smart homes integrated with automated technologies for elderly care \protect\cite{Majumder2017SmartChallenges}}
%\label{fig_sim3}
%\end{figure}

\section{User Perceptions}
\label{sec:perspective}

Environmental Gerontology (EG) adapts technology to improve layouts and set-ups of home environments, enabling elderly to perform their daily tasks with ease \cite{Carnemolla2018AgeingIntersect}. This influences the emotional and behavioural functioning of the older person, specifically their independence and well-being, where a dependency on care can lead to a loss of dignity. Tailored smart home environments technology supports `ageing in place' where elderly maintain autonomy and independence while their health conditions are passively monitored reducing institutional care and easing the economic burden on health \cite{FahimDailySmartphone, Carnemolla2018AgeingIntersect}. 

Currently, elderly seem reluctant to engage with smart homes systems. This is influenced by gender, age, voluntariness of use and prior experience \cite{Pal2018Internet-of-ThingsPerspective}. Carnemolla \cite{Carnemolla2018AgeingIntersect} developed an EG model to identify the role of technology in ageing in place and performed three case studies on elderlys experiences in receiving care and living in smart homes \cite{Carnemolla2018AgeingIntersect}:
\begin{enumerate}
\item Case Study 1 examined the use of automatic light sensors in the home of a 90-year old woman living independently, suffering from arthritis and gastro-oesophageal reflux disease. The woman gained confidence to visit the bathroom at night without needing to search for light switches. 
\item Case Study 2 examined the use of video doorbell locks in the house of a 79-year old woman living independently and being treated for breast cancer and diabetes. Her mobility was limited and was experiencing anxiety due to regular door knocks by strangers. The triggering of the video doorbell and lock system was screened by her informal caregivers and her anxiety was reduced. However, her inability to understand this technology limited its use. 
\item Case Study 3 examined the effect of using robotic lawn mowers for an 82-year old man living with his wife and suffering from Chronic Obstructive Pulmonary Disease (COPD) and limited mobility. The robotic mower provided a sense of independence and restored his dignity, improving his well-being. 
\end{enumerate}

\section{Advantages of smart homes for elderly}
\label{sec:advantages}

Smart homes supports well-being for elderly living independently in their homes \cite{Carnemolla2018AgeingIntersect} through: 
\begin{enumerate}
    \item Technology providing the elderly support and assistance to live independently in their homes, improving the quality of life \cite{Wang2014AnNetworks}. 
    \item Entertainment and social interactions for the elderly at home. 
    \item Notifications of early predictions and management of health risks from continuous and real-time monitoring of physiological signs and behavioural patterns.
    \item Accurate and reliable data collected to improve the quality and management of care and services assisting formal carers and health professionals in making better decisions and timely intervention by providing early warnings of health deteriorating.
    \item Informing and educating the elderly to self-manage their health empowering them with positive attitudes while reducing the frequency of hospital visits and length of stay. 
\end{enumerate}

Despite these advantages, smart homes continue to have challenges. 

\section{Research challenges for smart homes}
\label{sec:challenges}

The adoption rate of smart home technologies is low due to \cite{HolzingerHuman-ComputerBeyond, liu2016smart, arthanat2019profiles, li2021motivations}:
\begin{enumerate}
    \item Data security and privacy issues deter elderly from adapting smart home systems, as they are reluctant to share their private health data. This lack of trust results in withheld information, misleading information, or avoidance of the system. 
    \item Limited studies on smart home usability, feasibility and viability.
    \item Fear of technology, doubting its usefulness where the effort required to learn new technology outweighs the perceived benefit.
    \item Technology anxiety from being accustomed to one technology and needing to learn another where the acceptance of new technology is relative to its ease of use.
    \item Smart home costs.
    \item Uninformative data.
\end{enumerate}

% An example of a floating figure using the graphicx package.
% Note that \label must occur AFTER (or within) \caption.
% For figures, \caption should occur after the \includegraphics.
% Note that IEEEtran v1.7 and later has special internal code that
% is designed to preserve the operation of \label within \caption
% even when the captionsoff option is in effect. However, because
% of issues like this, it may be the safest practice to put all your
% \label just after \caption rather than within \caption{}.
%
% Reminder: the "draftcls" or "draftclsnofoot", not "draft", class
% option should be used if it is desired that the figures are to be
% displayed while in draft mode.

\section{Smart Home Goal Feature Model}
\label{sec:shgfm}
We propose the Smart Home Goal Feature Model (SHGFM) to assist healthcare researchers and aged care industries design, develop and deploy smart homes for elderly. It emphasizes the need for researchers and industries to (1) assess the needs of the older residents, (2) establish measurable goals specific to those needs in (1), and (3) identify the features needed to meet those assessed needs and goals in (2)-(3). We present (i) the goals and (ii) the feature maturity layers, then (iii) map the goals to the maturity layers to derive the SHGFM.

\subsection{Goals} \label{sec:goals}

We adapted a heuristic approach to develop 7 levels. Each level has a unique measurable goal starting at Level 0 with no goals and no smart features through to Level 6 where systems provide monitoring and intervention delivery with emotional support. The defined goals are:

\begin{itemize}
\item Level 0: No goals
\item Level 1: Improving quality of life
\item Level 2: Preventative adverse health outcomes 
\item Level 3: Remote monitoring
\item Level 4: Improving response time of assessment and delivery of interventions
\item Level 5: Automated 'service of care' 
\item Level 6: Automated emotional support 
\end{itemize}

\subsection{Feature Maturity Layers}
\label{sec:maturity model}
We then used a heuristic approach to define features for each layer. Layer 0 is a home with no smart assistance. Layer 6 is an ideal fully automated layer to detect anomalies, select appropriate interventions and emotionally support elderly. This layer provides personalised rational decisions, emotional intelligence and companionship. Limitations in technology hinder the development of this Level. 

%The FML with its different layers has been shown in figure 4.\\

%\begin{figure} 
%\centering
%\includegraphics{FML_V3.png}
% where an .eps filename suffix will be assumed under latex, 
% and a .pdf suffix will be assumed for pdflatex; or what has been declared
% via \DeclareGraphicsExtensions.
%\centering
%\caption{The Feature Maturity Layers (FML)  with 7 different layers of features of smart homes}
%\label{fig_sim6}
%\end{figure}
\begin{itemize}
\item Layer 0: No Smart Assistance
%\subsubsection{Level 0: No Smart Assistance}

Layer 0 does not incorporate any smart devices or assistance in the home.

The features of Layer 0 include:
\begin{itemize}
\item No log of events
\item Manual operation of apparatus such as light switches, appliances and phones
\item No feedback loop
\end{itemize}
\bigskip
\item Layer 1: Stand-alone smart assistance
%\subsubsection{Level 1: Stand-alone smart assistance}

Layer 1 includes smart devices and systems supporting self-feedback where the elderly receives direct feedback from the smart home. For example, a smart home with smart lights turning on and off based on motion \cite{Carnemolla2018AgeingIntersect}. By integrating smart devices to other smart devices (including a smart hub or smartphone), it alerts the elderly of people at the front door or if a window is open \cite{Carnemolla2018AgeingIntersect}. Anomaly behaviour can be monitored and used to remind elderly to have a meal or take their medication \cite{FahimDailySmartphone, Moshnyaga2017ADementia}. 

The features of Layer 1 include: 
\begin{itemize}
\item Log of events
\item Automated sensors to trigger apparatus such as lights and appliances
\item Voice activated calls
\item Supports self-feedback loop (user receives direct feedback)
\item Self-feedback for anomaly behaviour
\end{itemize}
\bigskip
\item Layer 2: Alerts for anomaly detection
%\subsubsection{Level 2: Alerts for anomaly detection}

This layer includes smart devices and systems integrated with smartphones or tablets of elderly and informal and formal caregivers. The smart home provides the ability for caregivers to view the activities of elderly at any time and alert them of any anomaly behaviour. The alert can be sent via an app, SMS or email reminding the caregivers to check in on the elderly. %\autoref{fig_sim4} and \autoref{fig_sim5} 
\autoref{fig_sim4_sim5} presents a monitoring system capturing ``daily living activities'' from door contact and motion detection sensors placed in a home (\autoref{fig_sim4}) (\autoref{fig_sim5}) and alerts informal caregivers on an anomaly detection. \autoref{fig_sim5} is based on \cite{Lotfi2012SmartBehaviour}. However, it has been modified to remove the signals from the corridor from the time series data as it is not an endpoint, specifically the person does not rest in the corridor as they are walking to a specific endpoint room, such as a bedroom or the kitchen. Billy is an off-the-shelf device that uses motion sensors to detect the opening of doors or boxes, such as the front door, refrigerator door and medication box. This data is used to monitor and learn elderly's daily routines from patterns identified in eating and sleeping, medication management, entering and leaving the house and movements within the house \cite{PartnersofECH2021WhatBillyb}. A change in behaviour triggers an alert to informal caregivers registered in the system. HomeGuardian is an assistive technology that learns about every object in the room and the behaviour of the elderly through their interactions with the system \cite{HomeGuardianAIPtyLtd2021HomeGuardian}. It monitors the movement of the users, detecting falls and irregular incidents, and alerting formal caregivers without relying on human activated emergency buttons. The system also ensures privacy and security by processing data on the device, increasing the trust of the users on the system. Sofihub monitors elderly's health and well-being by learning their behaviour and sending medication reminder notifications, event reminders, morning greetings, weather updates, and bed reminders \cite{2021SOFIHUB}. ARMED uses wearable technology to prevent fall incidents by identifying risks up to 32 days in advance \cite{2021HASTechnology}. It uses predictive analytics and machine learning to enable early intervention by clinicians. It performs remote risk monitoring of health conditions by identifying daily escalating risk, escalating risk trends over a 14-day period and if the user is over or under exerting themselves. Formal caregivers are alerted to provide necessary care prior to an incident occurrence. 

The added features of Layer 2 include:
\begin{itemize}
\item External-feedback loop to carers
\item External-feedback for anomaly behaviour
\end{itemize}
\bigskip
%\begin{figure}
%\centering
%\includegraphics{Figure4.png}
%\centering
%\caption{An example of the layout of the house and the location of door contact sensor and motion detection sensor \cite{Lotfi2012SmartBehaviour}}
%\label{fig_sim4}
%\end{figure}

%\begin{figure}
%\centering
%\includegraphics{Figure5.png}
%\centering
%\caption{Monitoring of daily activities inside the house with respect to time by the smart sensors shown in \autoref{fig_sim4} \cite{Lotfi2012SmartBehaviour}. The blue signal represents daily activity in each room as time series data, enabling prediction of future activity patterns of the resident.}
%\label{fig_sim5}
%\end{figure}

\begin{figure}[htb]
     \centering
     \begin{subfigure}{\linewidth}
         \centering
         \includegraphics[width=0.9\linewidth]{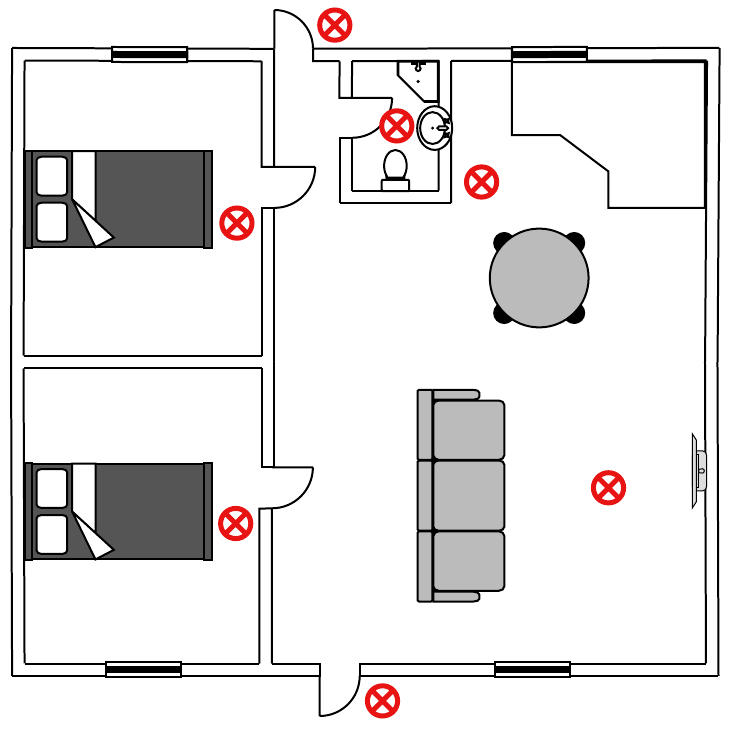}
         \caption{An example of the layout of a house at a residential aged care facility and the location of door contact sensors and motion detection sensors.}
         \label{fig_sim4}
     \end{subfigure}
     \begin{subfigure}{\linewidth}
         \centering
         \includegraphics[width=0.9\linewidth]{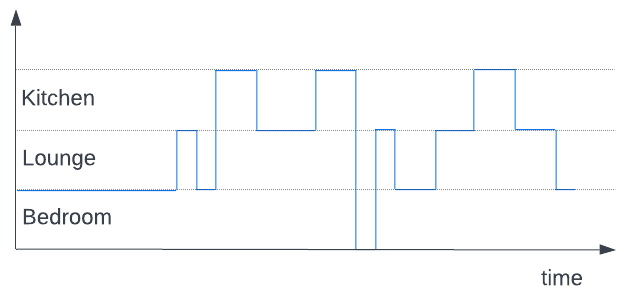}
         \caption{Monitoring of activities relative to time using sensors modified from \cite{Lotfi2012SmartBehaviour}.  The signal represents activity in each room as time series data, enabling prediction based on patterns of the resident.}
         \label{fig_sim5}
     \end{subfigure}
        \caption{An example layout of the house with sensors located and monitoring of daily activities by the smart sensors}
        \label{fig_sim4_sim5}
\end{figure}

\item Layer 3: Virtual Health Assistant
%\subsubsection{Level 3: Virtual Health Assistant}

This layer includes devices and systems supporting `standard of care' services from health care facilities, including hospitals and residential aged care facilities, where formal caregivers regularly monitor real-time data streams. While Layer 2 provides alerts post detection of any anomaly behaviour, Layer 3 includes the `standard of care' services offered by care facilities providing preventative care by formal caregivers. This increases the care provided during the day when formal caregivers can determine the onset of any anomaly behaviour and intervene before an incident occurs. For example, the  BioBeats Early Warning Score (EWS) System \cite{2021BiobeatSmartmonitoring} provides a glanceable user interface for health professionals to monitor the health status of elderly for early detection of health deterioration acting as a remote health assistant. This layer does not include automated interventions. 

The added features of Layer 3 include:
\begin{itemize}
\item Continuous remote monitoring of real-time data by clinicians
\item `Standard of care' services for preventative care. 
\end{itemize}
\bigskip
\item Layer 4: Clinically validated smart assistance

Layer 4 integrates smart devices with clinical systems, providing risk analysis algorithms for system assessment and risk stratification. Human intervention is required, where clinicians accept or override system assessments with clinical assessments. This provides a feedback loop where the algorithm for the risk stratification can be improved based on the discrepancies between the system and clinical assessments. CAPRI RPMS (Cancérologie Parcours Région Ile de France Remote Patient Monitoring Systems) \cite{Ferrua2020HowStudy} and PiMS \cite{logothetis2022pims} are systems that include interventions that are activated by the humans-in-the-loop and includes calling the elderly to follow up and/or connecting to an ambulance service. A call centre is also available to support the older person. The recommended interventions can be linked to the risk stratification layer. 

\begin{figure*} 
\centering
\includegraphics{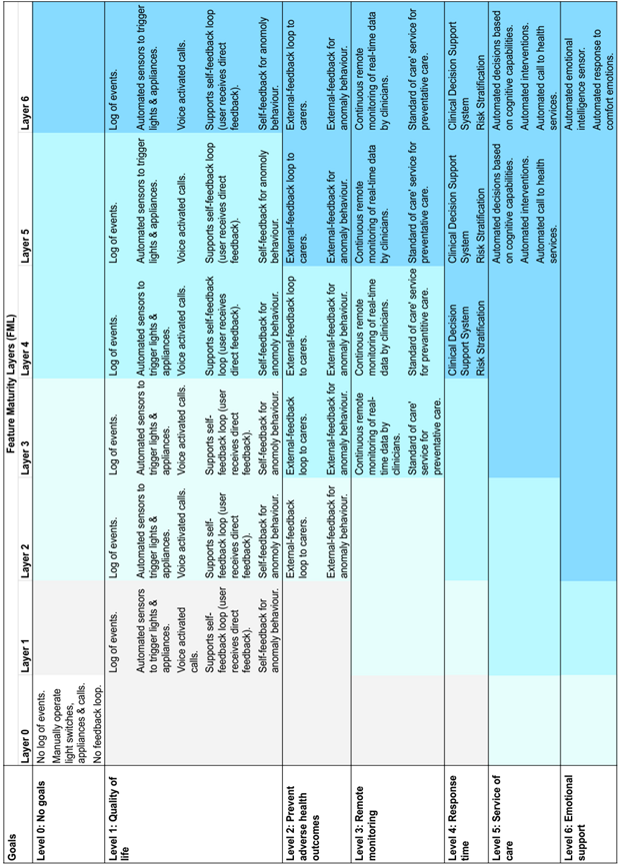}
\caption{The Smart Home Goal Feature Model with 7 different levels of goals and layers of features from Feature Maturity Layers (FML)}
\label{fig_sim7}
\end{figure*}

\begin{figure*}[hbt!]
\centering
\includegraphics[width=0.9\linewidth]{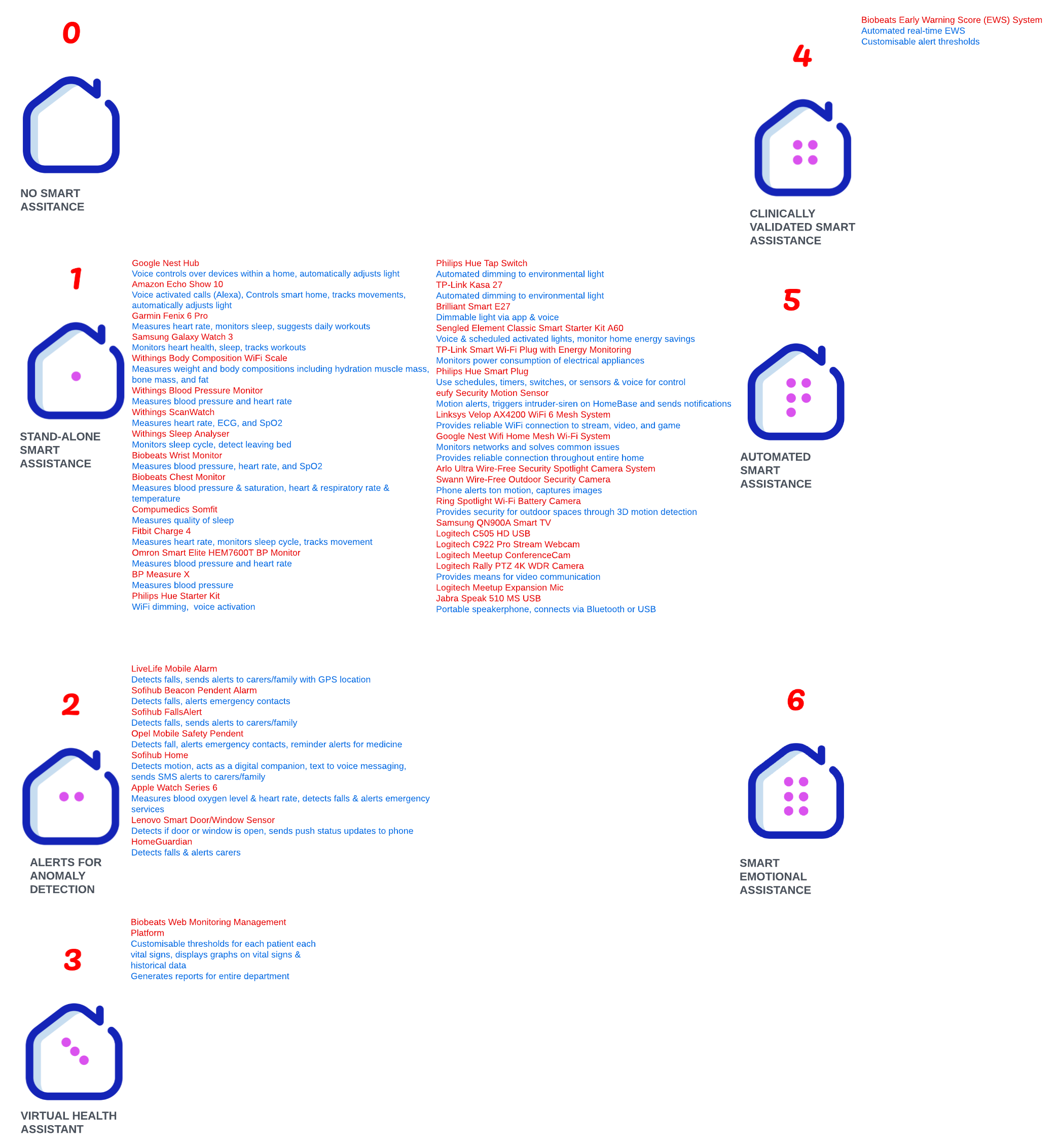}
\caption{Smart homes devices mapped to feature maturity layers}
\label{fig_sim3}
\end{figure*}

The added features of Layer 4 inlcude:
\begin{itemize}
\item `Standard of care' services for preventative care
\item Clinical Decision Support System
\item Risk stratification
\end{itemize}
\bigskip
\item Layer 5: Automated smart assistance
%\subsubsection{Level 5: Automated smart assistance}

The concept of Layer 5 is to remove the humans-in-the-loop and provide smart homes that are fully automated and can recommend appropriate and tailored interventions. The algorithms automate decisions based on cognitive capabilities (by collecting signals from the brain and muscles). The smart devices would deduce the status of elderly by measuring and monitoring their body signals, movements, voice and disease biomarkers. Any abnormal behaviour would trigger automated decisions about the future course of activities and action the required interventions. In case of extreme abnormal behaviour, it would alert emergency services. Currently, this has been discussed at a conceptual level \cite{Alhussein2018CognitiveMonitoring} with the potential of a prototype in the near future.

The added features of Layer 5 include:
\begin{itemize}
\item Automated decisions based on cognitive capabilities
\item Automated interventions
\item Automated calls to health services
\end{itemize}
\bigskip
\item Layer 6: Smart emotional assistance

This layer offers complete automation, adding emotional intelligence to Layer 5s self reasoning. The emotional intelligence of the smart home can detect the mood of elderly to provide emotional comfort. For example, a sense of loneliness can trigger a conversation with an avatar or to play calming music. 

The added features of Layer 6 include:
\begin{itemize}
\item Automated emotional intelligence sensor
\item Automated comfort to emotions
\end{itemize}

\end{itemize}

A market research analysis was conducted and common devices identified to support ageing in place were mapped to the feature maturity layers in \autoref{fig_sim3}. 

\subsection{Goal Feature mapping - Smart Home Goal Feature Model (SHGFM)}
\label{sec:gfm}

The SHGFM is derived by mapping the feature maturity layers of a smart home to goals from our literature review. However, the needs of elderly can be subjective and needs to be assessed by SMEs case by case to map to the Smart Home level. This is outside the scope of this work, we are providing a model for SMEs to use. Based on assessed needs, goals can be established at each level described in \autoref{sec:goals}, and the number of features for each level chosen from the Feature Maturity Layers (FML) described in \autoref{sec:maturity model}. 

Refer to \autoref{fig_sim7} for the detailed SHGFM with the goals and corresponding features for each layer. To demonstrate the models use, let us consider an example of an older resident, Mr Gibson. His formal caregiver (SME) assesses Mr Gibsons needs and identifies that he (1) lives by himself, (2) does not have chronic health conditions, (3) can take care of himself, and (4) has one of his children visiting him daily as an informal caregiver. The formal caregivers assessments show that Mr Gibson can be put on the minimal established goal at `Level 1: Quality of life' and FML 1 can be identified. Mr Gibson and his formal caregiver feel assured that the features are consistent for the higher layers, so he can have an easy transition to the next layer if his health deteriorates, and he needs extra support. FML 1 has the features of a basic smart home with logging of events, automated sensors triggering lights and appliances, voice activated calls and self-feedback to the resident about the daily routine and anomaly behaviour. However, if Mr Gibson had a fall incident when he was alone at home, the minimal established goal can change to `Level 2: Prevent adverse health outcomes' and FML 2 can be provided incorporating the features of external feedback to informal caregivers on Mr Gibsons routine and any anomaly behaviour. Thus, based on the assessed needs, residents can transition to higher level goals with their corresponding features from FML. This model can assist healthcare researchers and aged care industries to set up smart homes for the elderly and support ageing in place.

\section{Conclusion and future perspectives}
\label{sec:conclusion}

This paper synthesises the knowledge in smart home technologies supporting ageing in place into a structured Smart Home Goal Feature Model (SHGFM) providing a tool that can be used by SMEs to identify the level of a smart home elderly need based on case assessments. A caregiver can (i) establish goals based on elderly's needs and (ii) design, develop and deploy that smart homes against the SHGFM features for the goal. 

Supporting 'ageing in place', smart home technologies can offer a quality of care relating to health and mental well-being through (i) environmental monitoring, (ii) physiological monitoring, (iii) monitoring motion and behaviour, and (iv) providing social interactions and entertainment. Previous research indicates current holistic approaches towards smart home systems for elderly results in low adoption. Understanding the needs of the older person and their dependence on human characteristics, social background and dependence on technology has not been considered, causing a mismatch between the users’ expectations and services available. Research shows that hesitation and conservative attitudes towards new technology is common in this generation of elderly. This hesitation can be due to poor interface design, issues of privacy and trust, economic and educational barriers.

For elderly, a key role in its acceptance is to have a strong intention and willingness to use new technology in addition to the perception of technology being beneficial. Smart homes for health support can enable elderly to manage their health with better access to healthcare facilities, resulting in an improved  quality of life. This can improve their perception towards smart homes. An increase in uptake of smart homes by elderly can be facilitated by (i) integrating support from clinical experts and (ii) supporting family members that initiate and coordinate the technologies in their loved one's homes. 

Smart homes require user-centered and age appropriate approaches. The challenge faced by healthcare researchers and aged care industries is the correct assessment of the needs of elderly and mapping their needs to the goals and features for a smart home. The proposed SHGFM attempts to resolve this issue by defining a set of goals at different levels mapped to a certain set of features. Our model aims to provide guidance to healthcare researchers and aged care industries to set up smart homes based on the needs of elderly as a case by case assessment. Evaluating smart homes in a diverse older population can significantly improve the quality of life of elderly, assist to monitor and control their health conditions and improve its acceptance.

% if have a single appendix:
%\appendix[Proof of the Zonklar Equations]
% or
%\appendix  % for no appendix heading
% do not use \section anymore after \appendix, only \section*
% is possibly needed

% use appendices with more than one appendix
% then use \section to start each appendix
% you must declare a \section before using any
% \subsection or using \label (\appendices by itself
% starts a section numbered zero.)
%

% \appendices
% \section{Proof of the First Zonklar Equation}
% Appendix one text goes here.

% % you can choose not to have a title for an appendix
% % if you want by leaving the argument blank
% \section{}
% Appendix two text goes here.
% \section*{Authors' contributions}
% Irini Logothetis: Conceptualization, Methodology, Formal analysis, Writing – original draft, Writing –  review \& editing, Priya Rani: Methodology, Formal analysis, Writing – review \& editing, Shangeetha Sivasothy: Formal analysis, Writing - review \& editing, Rajesh Vasa: Conceptualization, Writing – review \& editing, Kon Mouzakis: Conceptualization, Writing – review \& editing

% % use section* for acknowledgment
\section*{Acknowledgment}
 This research was conducted by the Australian Research Council Industrial Transformation Hub for Digital Enhanced Living (Project ID: IH170100013). The authors acknowledge the Australian Government and Uniting AgeWell Limited Victoria for funding and supporting this research.

\section*{Statement on conflicts of interest}
The authors declare that they have no known competing financial interests or personal relationships that could have appeared to influence the work reported in this paper.

% Can use something like this to put references on a page
% by themselves when using endfloat and the captionsoff option.
%\ifCLASSOPTIONcaptionsoff
%  \newpage
%\fi

% trigger a \newpage just before the given reference
% number - used to balance the columns on the last page
% adjust value as needed - may need to be readjusted if
% the document is modified later
%\IEEEtriggeratref{8}
% The "triggered" command can be changed if desired:
%\IEEEtriggercmd{\enlargethispage{-5in}}

% references section

% can use a bibliography generated by BibTeX as a .bbl file
% BibTeX documentation can be easily obtained at:
% http://mirror.ctan.org/biblio/bibtex/contrib/doc/
% The IEEEtran BibTeX style support page is at:
% http://www.michaelshell.org/tex/ieeetran/bibtex/
%\bibliographystyle{IEEEtran}
% argument is your BibTeX string definitions and bibliography database(s)
%\bibliography{IEEEabrv,../bib/paper}
%
% <OR> manually copy in the resultant .bbl file
% set second argument of \begin to the number of references
% (used to reserve space for the reference number labels box)

%\bibliographystyle{IEEEtran}
%bibliography{IEEEabrv,references_copy}
\bibliographystyle{elsarticle-harv}
\bibliography{references_copy}

% \begin{thebibliography}{1}

% \bibitem{IEEEhowto:kopka}
% H.~Kopka and P.~W. Daly, \emph{A Guide to \LaTeX}, 3rd~ed.\hskip 1em plus
%   0.5em minus 0.4em\relax Harlow, England: Addison-Wesley, 1999.

% \end{thebibliography}

% biography section
% 
% If you have an EPS/PDF photo (graphicx package needed) extra braces are
% needed around the contents of the optional argument to biography to prevent
% the LaTeX parser from getting confused when it sees the complicated
% \includegraphics command within an optional argument. (You could create
% your own custom macro containing the \includegraphics command to make things
% simpler here.)
%\begin{IEEEbiography}[{\includegraphics[width=1in,height=1.25in,clip,keepaspectratio]{mshell}}]{Michael Shell}
% or if you just want to reserve a space for a photo:

% \begin{IEEEbiography}{Michael Shell}
% Biography text here.
% \end{IEEEbiography}

% % if you will not have a photo at all:
% \begin{IEEEbiographynophoto}{John Doe}
% Biography text here.
% \end{IEEEbiographynophoto}

% % insert where needed to balance the two columns on the last page with
% % biographies
% %\newpage

% \begin{IEEEbiographynophoto}{Jane Doe}
% Biography text here.
% \end{IEEEbiographynophoto}

% You can push biographies down or up by placing
% a \vfill before or after them. The appropriate
% use of \vfill depends on what kind of text is
% on the last page and whether or not the columns
% are being equalized.

%\vfill

% Can be used to pull up biographies so that the bottom of the last one
% is flush with the other column.
%\enlargethispage{-5in}

% that's all folks
\end{document}